# Magnetic steganography based on wide-field diamond quantum microscopy

Jungbae Yoon[1,+], Jugyeong Jeong[1,+], Hyunjun Jang[1], Jinsu Jung[1], Yuhan Lee[1], Chulki Kim[2], Nojoon Myoung[3,4], & Donghun Lee[1,*]

[1]Department of Physics, Korea University, Seoul, 02841, Republic of Korea

[2]Quantum Technology Center, Korea Institute of Science and Technology (KIST), Seoul, 02792, Republic of Korea

[3]Department of Physics Education, Chosun University, Gwangju, 61452, Republic of Korea

[4]Institute of Well-Aging Medicare & Chosun University G-LAMP Project Group, Chosun University, Gwangju 61452, Republic of Korea

[+] These authors contributed equally to this work.

[*]Corresponding authors.
Email addresses: donghun@korea.ac.kr

## Abstract

We experimentally demonstrate magnetic steganography using wide-field quantum microscopy based on diamond nitrogen-vacancy (NV) centers. The method offers magnetic imaging capable of revealing concealed information otherwise invisible with conventional optical measurements. For a proof-of-principle demonstration of the magnetic steganography, micrometer structures designed as pixel arts, barcodes, and QR codes are fabricated using mixtures of magnetic and non-magnetic materials: Ni and Au. We compare three different imaging modes based on the changes in frequency, linewidth, and contrast of the NV's electron spin resonance, and find that the last mode offers the best quality of reconstructing hidden magnetic images. By simultaneous driving of the NV's qutrit states with two independent microwave fields, we expedite the imaging time by a factor of three. This work shows potential applications of quantum magnetic imaging in the field of image steganography.

## Introduction

Steganography, originating in the fifth century BC, is a longstanding technique for concealing information[1]. In ancient Greece, a clandestine message was inscribed on the shaved head of a slave, which was subsequently covered by the growth of hair. Since that time, steganography has made significant progress and has emerged as a potent tool in the era of digital technology for securing information[2,3]. Unlike cryptography, its primary objective is to maintain the confidentiality of original digital content by overlaying it with another file, therefore enabling the embedded text, image, audio, or video files to be delivered with minimal attention[2,3]. For instance, image steganography is employed to conceal medical record images that include valuable medical history data and personal information of the patient[4].

Magnetic steganography is a widely used technique in the fields of finance, digital security, and forensic science[5-7]. Magnetic structures patterned by ink containing magnetic nanoparticles are typically employed for the purpose of examining currency, counterfeit bills, and secure documents[8]. For the detection of magnetic signals, magnetic devices such as magneto-inductive sensors and giant magneto resistance (GMR) sensors are commonly used, while, for high-resolution images, scanning magnetometers such as magnetic force microscope (MFM), scanning Hall probe microscope, superconducting quantum interference device (SQUID) microscope, and scanning diamond nitrogen-vacancy (NV) center are employed[9-16]. However, these measurements either require large magnetic objects or are limited by a smaller scan size and long imaging time. Therefore, a novel imaging technique enabling detection of small objects over a large area with fast measurement time will be crucial for the advance of steganography.

This work presents a novel approach to magnetic steganography using wide-field quantum microscopy. The diamond NV center is a solid-state qubit possessing high magnetic field sensitivity and spatial resolution[13-17]. Owing to the unique advantages, it has become one of the primary qubit platforms used in the field of quantum sensing and imaging[13-22]. Wide-field quantum microscopy provides magnetic imaging over millimeter-sized areas with hundreds of nanometers spatial resolution and second-order imaging time[18-22].

In this paper, the wide-field quantum microscope is used to image concealed magnetic structures and obtain secure information. Micrometer magnetic structures are fabricated with mixed patterns made of magnetic and non-magnetic materials: Ni and Au. Optical images provide limited identification of the two materials, whereas magnetic images exclusively show magnetic parts that are made of Ni, therefore revealing hidden magnetic information. For a proof-of-principle experiment, we fabricated a heart-shaped pixel art structure, in which a diamond-shaped magnetic pattern is integrated. Using the wide-field quantum microscope, we are able to visualize the concealed feature that is invisible in optical images. Beside the pixel art image, we also fabricated magnetic structures with hidden information: one-dimensional barcodes and two-dimensional quick-response (QR) codes. We show that optically indistinguishable information was revealed from magnetic images. We also compared three imaging modes possible in the wide-field quantum microscopy measurement: mapping the change in frequency, linewidth, and contrast of the NV's electron spin resonance (ESR), and found that the last mode is more effective than others suitable for the steganography applications. Finally, we demonstrated a dual driving method that utilizes the NV's qutrit states instead of qubit states. This method significantly enhances sensitivity and hence reduces imaging time by a factor of three. This work introduces a novel approach to implementing quantum imaging techniques in the field of steganography.

## Results and discussion

### Steganography with magnetic pixel arts

The concept of magnetic steganography based on wide-field quantum microscopy is illustrated in Fig. 1(a). A fabricated sample with a magnetic hidden structure is placed on a single crystal diamond plate (2 x 2 x 0.1 mm$^3$; [$\bar{1}$00] top face). An ensemble of NV centers, with a density of ~ 10 ppm, is uniformly distributed over the plate at a depth of ~ 15 nm from the diamond's top surface. An Omega-shaped gold microwave waveguide patterned on a cover glass and located underneath the diamond plate generates a microwave field at ~ 2.9 GHz to drive the NV's spin transitions[19,20]. A green laser ($\lambda$ = 532 nm; power = 700 mW) is used to excite the NV centers within a field-of-view, e.g. ~ 100 μm$^2$, of a total internal reflection fluorescence (TIRF) objective lens (N.A. = 1.49). The NV's fluorescence is collected by the objective lens and imaged by a scientific complementary metal–oxide–semiconductor (sCMOS) camera. The NV's spin-dependent fluorescence, driven by microwave radiation around 2.9 GHz, facilitates optically detected magnetic resonance (ODMR) measurements[17], yielding magnetic images in this study. For optical images of the samples, on the other hand, we employed a red light-emitting diode (LED; $\lambda$ = 625 nm, power = 500 mW), with the transmission post-sample being captured by the sCMOS camera.

At the ground state, the negatively charged NV center (NV⁻; referred to as NV throughout the paper) exhibits a spin triplet (S = 1) configuration[17]. As seen in Fig. 1(b), the ODMR measurement shows the NV's photoluminescence specific to the spin resonances occurring at frequencies of $m_s = 0 \leftrightarrow m_s = -1$, and $m_s = 0 \leftrightarrow m_s = +1$. The degenerated resonances can be split by external magnetic field, $B_{ext}$. The difference in the transition frequencies corresponds to the Zeeman splitting, which allows us to determine the magnitude of static magnetic field along the NV's quantization axis. Magnetic field image is generated by mapping the Zeeman splitting at each pixel in the fluorescence map captured by the camera. More details about the wide-field quantum microscope setup and the ODMR measurement are available in the Methods and Supplemental Materials.

In order to demonstrate the idea of magnetic steganography, we fabricated a heart-shaped pixel art composed of square dots, which are 1 x 1 μm² in size and are spaced 2 μm apart. The dots consist of either magnetic or non-magnetic materials, i.e., Ni and Au. The magnetic dots are interspersed in a diamond pattern, which serves as the concealed image in this sample. As shown in Fig. 1(c), the optical image barely reveals the hidden structure, but the magnetic field map clearly displays the diamond pattern. The test measurement indicates that magnetically encoded information can be safely protected against conventional optical measurement. Furthermore, deciphering the information requires a magnetic imaging device capable of covering large areas of several hundreds of μm² with a spatial resolution of less than 1 μm.

Steganography with magnetic barcodes and QR codes

For a proof-of-principle demonstration of encrypting magnetic information, we fabricated two steganography samples, barcodes and QR codes, as depicted in Fig. 2. First, barcodes, also known as universal product codes (UPCs) can encode both numerical values and alphabetic characters by combining rectangle bars of varying widths and spacings[23]. For each number or character, the combined length of the width and spacing should remain constant and, if one bar increases in width, for example, the spacing must be reduced by the same amount, so ensuring a uniform interval for all numbers and characters included in the barcode.

We constructed barcode structures comprising Ni and Au rectangle bars with 20 μm in length. These bars were designed with two distinct widths and spacings: 2 μm and 4 μm. The numbers and characters of the barcode are determined by their relative position, width, and spacing. We modified the code 39, commonly used for optical barcodes[24], by incorporating a magnetic degree of freedom and established a new table of barcode characters, Table S1, in the Supplemental Materials. Utilizing the table, we created two distinct magnetic barcodes that encode the characters "NV" or "KR". Figure 2(a) shows the optical and magnetic field images of the barcodes, together with the assigned character codes. The optically identical barcodes display distinct magnetic images corresponding to the designated characters. Note that we included Null characters both before and after the message characters in order to designate the start and end of the message. Besides the steganographic capability, magnetic barcodes offer an additional advantage over optically detected barcodes. Every magnetic bar affords an extra degree of freedom, whether with or without a magnetic field, leading to a more compact pattern with fewer bar configurations in comparison with solely non-magnetic structures used for optical readout.

We extend the method employed in the barcode measurements to two-dimensional matrix barcodes, i.e., QR codes. Together with barcodes, QR codes are commonly used in our daily life for methods of identifying and tracking products and websites in markets and mobile phones[25]. The data capacity of a QR code is determined by both the quantity of symbols, i.e., black dots, and the degree of error correction. QR codes function according to the Reed-Solomon code, which consists of four error correction levels: L (low), M (medium), Q (quartile), and H (high)[26]. The higher level allows for more accurate code recognition, but it also results in a reduced amount of data being stored due to the increased use of symbols for error correction. On the other hand, a QR code classified as level L requires a minimum of 93% of the symbols to be accurately arranged in order to correctly reconstruct the original information.

We constructed 25 x 25 matrix QR codes by using Ni and Au square dots with dimension of 2 x 2 $\mu m^2$ and a spacing of 1 $\mu m$. Figure 2(b) displays the optical and magnetic field images obtained from two distinct magnetic QR codes. The optically indistinguishable codes are clearly recognized in the magnetic images as encoded website links to the homepages of Korea University (http://www.korea.ac.kr) or the authors' laboratory (http://www.qdl.korea.ac.kr). The results obtained from the experiments on barcodes and QR codes successfully demonstrate the capability of wide-field quantum microscopy in securing information contained within magnetic images.

## Magnetic imaging modes and dual driving method

The magnetic field image is one of the three imaging modes offered by wide-field quantum microscopy. Figure 3(a) displays an example of the ODMR spectrum measured from the ensemble NV centers whose photoluminescence are plotted as a function of microwave frequency. The photoluminescence is changed at the frequencies of the NV's spin resonances, $m_s = 0 \leftrightarrow m_s = -1$ and $m_s = 0 \leftrightarrow m_s = +1$, appearing as dips or negative peaks in the plot. Our ensemble measurements reveal four pairs of ODMR resonances, corresponding to the four distinct groups of NV centers aligned with the crystal axes $[111]$, $[\bar{1}\bar{1}1]$, $[1\bar{1}\bar{1}]$, and $[\bar{1}1\bar{1}]$. We applied an external magnetic field of ~ 400 G to separate the pairs, and only one pair of the resonances, e.g., $[111]$, was utilized in our imaging experiments.

We acquired magnetic field images by measuring the amount of the Zeeman shift in the resonance frequencies or their relative splitting. In addition to the Zeeman shift, the linewidth and height (also known as contrast) of the resonance can also serve as suitable imaging modes. Near magnetic structures, the presence of non-zero magnetic noise or field components that are perpendicular to the quantization axis of the NV center can adversely affect the coherence time of the NV or result in the spin states to mix, leading to variations in the linewidth and contrast[20,27,28]. The wide-field imaging experiments involve the simultaneous measurement of frequency, linewidth, and contrast in each pixel, so enabling the acquisition of three distinct imaging modes from a single measurement.

Figure 3(b) compares the imaging modes used for the samples of pixel arts, barcodes, and QR codes. In wide-field quantum microscopy research, the imaging mode based on frequency shift is often employed to obtain quantitative data on the B field of the magnetic samples being studied[18-22]. On the other hand, the other modes illustrate relative variations in the images and thus offer solely qualitative data about the samples. Nevertheless, this is sufficient for the most magnetic steganography applications, which aim to reveal hidden magnetic structures rather than their detail magnetic properties, and the concealed information is generally digital, either with or without a magnetic field. Moreover, the frequency shift or B field imaging technique may distort the image of original structures and lead to a misleading interpretation, particularly when the dots or building blocks of the image are closely packed and thus the fields between them can be cancelled out. We tested this effect using magnetic simulations based on the modes.

Figure 4 presents a comparison of three imaging modes based on micromagnetic simulation on square dots, that are 1 x 1 $\mu m^2$ in size and separated by 2 $\mu$m. An opensource software called the object oriented micromagnetic framework (OOMMF)[29] is used to simulate the three-dimensional profiles of magnetic stray field around a sample with a mesh size of 50×50×50 $nm^3$. We project the stray field onto the NV's quantization axis in every pixel of the image to replicate the ODMR data and generate the final simulation results of frequency shift, linewidth and contrast changes. First, we conducted micromagnetic simulation as well as scanning magnetometry measurement based on a single NV center, as depicted in Figure 4 (a) and (b). The scanning magnetometry has been used to investigate magnetic materials and current transport devices in order to obtain magnetic imaging with high spatial resolution on the order of 10 nm[13-15]. For both the simulation and experiment, the NV center is positioned approximately 1 µm above the sample. Depending on whether the B field is entering or exiting from the Ni dot, the images exhibit opposite signs of B field. This situation can give rise to complications when the two dots are closely positioned and their distance is smaller or comparable to the field profile, resulting in the overlap and cancellation of the field at the areas between the dots. This effect becomes more pronounced when we employ wide-field quantum microscopy.

In contrast to single spin scanning magnetometry, wide-field quantum microscopy utilizes multiple NV centers within a diffraction-limited laser spot with a diameter of ~ 1 µm. This allows the measured ODMR data to be averaged out, resulting in the field averaging effect[20,27,28]. The effect not only decreases the total magnitude but also leads to a more extensive profile of the B field. Figure 4(c) depicts micromagnetic simulation that replicates the experimental conditions, including the field averaging effect and the experimental configurations of our wide-field quantum microscope. Closely spaced dots result in a further reduction of the overall B field due to the field averaging effect and the field overlap between the dots. Furthermore, both the frequency shifts and linewidth variations are more pronounced around the edges of the dots, whereas the contrast images display more concentric profiles in the images. From the comparison of both the simulation and experiment, we conclude that the contrast imaging mode provides better imaging quality than the others and is more appropriate for magnetic steganography applications. Note that the underlying mechanism of the contrast image differs from that of a single NV center measurement, which mainly results from magnetic noise or fluorescence suppression due to the spin mixing[30, 31]. In this study, however, the field averaging effect is the primary source of the changes in the contrast. More details about the micromagnetic simulation and field averaging effect are available in the Methods and Supplemental Materials.

Finally, we developed an expedited imaging method by employing the dual driving scheme [32-34]. In its ground energy states, the NV center is composed of three spin levels. Qubit state based on either $m_s = 0$ and $m_s = -1$ or $m_s = 0$ and $m_s = +1$, is commonly used in quantum sensing and imaging experiments. By employing two microwave sources, it is possible to concurrently induce both transitions, achieving double quantum[32,33] or dual driving schemes[34]. Measurements based on the NV's qutrit states have been conducted with both pulsed and continuous wave (CW) microwave fields[32-34]. To the best of our knowledge, the latter method has not been applied in the wide-field quantum microscope experiment. Figure 5 presents a comparison of the contrast images on QR codes acquired by single driving with either $m_s = 0$ and $m_s = -1$ or $m_s = 0$ and $m_s = +1$, as well as dual driving with both transitions. Here we employed the Pearson correlation coefficient for quantitative comparison between single and dual driving, which is a statistical metric used to evaluate image quality[35].

The Pearson correlation coefficient between two image A, B is calculated by following equation.

$$P(A,B) = \frac{\sum_m \sum_n (A_{m,n} - \bar{A})(B_{m,n} - \bar{B})}{\sqrt{\{\sum_m \sum_n (A_{m,n} - \bar{A})^2\}\{\sum_m \sum_n (B_{m,n} - \bar{B})^2\}}}$$

, where $\bar{A}$ and $\bar{B}$ stand for the mean values of image A and B.

It measures the correlation between an image taken at a certain instant and the final image obtained after a prolonged period of measurement. The coefficient approaching 1 indicates that the image reaches the desired level of quality comparable to the final result.

In Figure 5(b), the Pearson correlation coefficient is plotted against the measurement (or imaging) time for two distinct cases: single driving and dual driving. In both cases, the coefficient approaches 1 as the measurement time increases. Under shorter time, the latter scenario exhibits a higher coefficient compared to the former, and the discrepancy increases as the measurement time decreases. For example, we compare their contrast images at two distinct measurement time: 100 and 1,000 seconds. As seen in Fig. 5(c), there exists a greater difference in the quality of the image at 100 seconds compared to 1,000 seconds. Achieving a coefficient level of ~ 0.65, which corresponds to the error correction level H, will require approximately three times more measurement time when using single driving. Therefore, the dual driving technique can substantially enhance the imaging speed in magnetic steganography applications. Note that the presences of noisy features appearing around the edges and corners of the QR images can be attributed mostly to the lower laser and microwave power in these regions close to the boundary of the microscope's field-of-view.

The improved quality in the dual driving can be understood by population changes in the NV's spin states. In the ODMR measurement, the contrast can be influenced by the relative population between $m_s = 0$ and $m_s = -1$ (or $m_s = +1$): a higher population at $m_s = 0$ results in lower contrast. By driving qutrit states simultaneously, the population in $m_s = 0$ is significantly reduced, while it substantially increases in $m_s = -1$ and $m_s = +1$. Consequently, there is an overall increase in the ODMR contrast. Figure 6 shows the calculated ODMR plots obtained from the numerical calculation on the NV population using rate equations. To solve the rate equations, we take into account a total of 7 states: 3 ground states and 3 excited states for spin triplets, and 1 shelving state for spin singlet. As seen in Fig. 6(a), we consider the excitation and relaxation rates among these states, i.e., $\Gamma_1$, $\Gamma_4$, $\Gamma_5$, $\Gamma_6$, and $\Gamma_p$, as well as the Rabi frequencies used for the spin driving, i.e., $\Omega_R$. Note that the hypothetical state 8 is necessary for the calculation so that the electrons at the ground states $1 - 3$ could be excited to the state 8 with the rate, $\Gamma_1$, and subsequently relaxed to the excited states $4 - 6$ with the rate, $\Gamma_4$. Adopted from Ref.36, we assigned the parameters as $\Gamma_1 = 0.61$ MHz, $\Gamma_4 = 50$ MHz, $\Gamma_5 = 3.0$ MHz, $\Gamma_6 = 0.25$ MHz, and $\Gamma_p = 4.87$ MHz.

Figure 6(b) compares the calculated ODMR results obtained from single and dual driving methods at three different microwave powers or Rabi frequencies, $\Omega_R$. Note that $\Omega_R$ is normalized by the Rabi frequency, $\Omega_R = 5.0 \times 10^6$ rad/s, that is estimated from the microwave power used in our imaging experiment. We also assume the same Rabi frequencies between $m_s = 0 \leftrightarrow m_s = -1$ and $m_s = 0 \leftrightarrow m_s = +1$ for simplicity. The case with antisymmetric Rabi driving is discussed in the Supplemental Materials. The calculation reveals that the ODMR contrast indeed increases when dual driving is implemented. In Fig. 6(c), we plot the calculated contrast, linewidth, and sensitivity as a function of $\Omega_R$. The sensitivity $S$ is determined from the contrast $C$ and linewidth $\Delta\omega$ after using the equation, $S \propto \Delta\omega/(C\sqrt{R_0})$, where $R_0$ represents the NV's fluorescence rate without applying the microwave fields. Although the linewidth difference between single and dual driving is minimal, the contrast and sensitivity plots clearly demonstrate that dual driving outperforms than single driving. We found approximately twice as much improvement in the optimal sensitivity around $\Omega_R = 1$ that is close to our experimental condition. Comprehensive information regarding the rate equations and theory calculations can be found in the Methods and the Supplemental Materials.

## Conclusion

This work presents a novel magnetic steganography technique that utilizes wide-field quantum microscopy. The method enables a scan area on the mm scale, a spatial resolution below μm, and imaging speeds on the order of minutes. Magnetic Ni patterns including pixel arts, barcodes, and QR codes are employed as concealed images, which are constructed in conjunction with non-magnetic Au layers. In contrast to traditional optical imaging, the magnetic imaging effectively reveals the hidden information. Encoded in magnetic barcodes and QR codes are information such as text messages and website links, indicating the potential of this approach in the field of image steganography. Our analysis revealed that the ODMR contrast image exhibits superior performance compared to the other two imaging modes, the ODMR frequency shift and linewidth mapping. Furthermore, we elucidated the averaging effect on the ODRM contrast by comparing it with the magnetic simulation. To expedite the imaging time, we implemented a CW dual driving technique that leverages the NV's qutrit states. This approach resulted in a threefold increase in the imaging time. Numerical calculation on the rate equations for the NV's 7 energy levels confirms our experimental findings.

Applications of the novel magnetic steganography in finance, forensic science, and digital security can encompass numerous fields. Here we demonstrated a proof-of-principle using a lithography technique to construct μm-size magnetic building blocks. Beyond the labor-intensive fabrication process described in this paper, we anticipate that more flexible and practical applications are possible with the use of advanced printing methods, such as piezo inkjet printing, drop impact printing and dip-pen nano-lithography[37-39].

## Methods

*Wide-field quantum microscope setup*

The custom-built wide-field quantum microscope setup comprises four primary components; (1) samples and sensors, (2) wide-field optical microscope, (3) microwave circuitry, and (4) control and data analysis. The methods for describing the magnetic samples and diamond sensors are outlined below. In the wide-field optical microscope, a green laser (wavelength = 532nm, optical power = 700 mW; Laser Quantum, Axiom 532) was employed, focused on the NV centers within a 100 x 100 μm² field of view using a total internal reflection fluorescence (TIRF) objective lens (N.V. = 1.49, magnification 60x; Nikon, CFI Apo TIRF 60XC Oil). The NV's fluorescence is collected by the objective lens and imaged by a scientific complementary metal–oxide–semiconductor (sCMOS; 2048 by 2048 pixels; 40 frames per second; Excelitas Technologies, PCO panda 4.2) camera after a 620-795 nm band-pass filter (Semrock, FF01-709/167). A red light-emitting diode (LED; wavelength = 625 nm, optical power = 500 mW) is used for optical imaging. For the microwave circuitry, we used a signal generator (Stanford Research Systems, SG384) to produce microwave fields around 2.9 GHz, which was subsequently amplified using a microwave amplifier (Mini Circuits, ZHL-25W-63+) and directed into an Omega-shaped gold waveguide fabricated on a cover glass. The synchronized operation of the laser, camera, and microwave circuit is managed by a custom-built software, while the acquired image data is processed by a workstation computer.

*Diamond sensors*

A commercial diamond plate (2 x 2 x 0.5 mm³; Element Six, electronic grade [$\bar{1}$00] top face) is sliced and thinned down to a thickness of 100 μm. NV centers are created by implanting $^{15}N^+$ into the diamond plate at 10 keV with a density of $10^{14}$ cm$^{-2}$ under vacuum, subsequently annealing the diamond at 1200 °C. Oxygen termination was accomplished by introducing Oxygen at a flow rate of 200 sccm at 465 °C to stabilize the charge state of the NV centers. The NV depth from the diamond surface is estimated to be 15 ± 5 nm.

*Magnetic steganography samples*

The magnetic and non-magnetic patterns were produced using electron beam lithography. First, a thin layer of electron-sensitive resist was coated on the cover glass. Electron beams were used to expose gold patterns in the resist followed by chemical treatments to develop the exposed areas. Subsequently, we evaporated a 10 nm titanium layer and a 50 nm gold layer with an electron beam evaporator. The titanium functions as an adhesive layer on the glass. Following the lift-off procedure, we obtained the gold patterns on the cover glass. We repeated the procedure for the nickel patterns, carefully aligning the electron beam exposure with the previously fabricated gold patterns.

*Magnetic simulation*

We utilized an opensource software known as the object oriented micromagnetic framework (OOMMF) to numerically compute the magnetic stray fields around magnetic samples with mesh sizes of $50\times50\times50$ nm$^3$. Subsequently, we projected the field along the NV axis to obtain simulated image measured by a single NV center. Even if we selected NV centers with the same crystal axis, in the wide-field measurements it is important to account for all ODMR resonances from the NV centers within the ~ 1 µm laser spot leading to an averaged magnetic field in each pixel. The final simulated magnetic images are derived after considering the field averaging effect, which accounts for our experiment conditions such as the quantity of NV centers in each pixel and the point spread function in the wide-field microscope configuration.

*Numerical calculation on double quantum scheme*

We performed a quantitative analysis to elucidate the mechanism underlying the enhanced sensitivity of the dual driving scheme. This analysis involves numerically solving coupled rate equations that characterize all transitions within the 7-level qutrit model of NV centers in a diamond. The solutions to these rate equations represent the populations of individual levels as functions of time, contingent upon system parameters such as microwave power or the Rabi frequency. After a period of time, the NV's photoluminescence (PL) rate is derived from 'steady-state' populations as follows:

$$R(\Omega_1, \Omega_2, \omega_1, \omega_2, s) = [\alpha n_{11} + \beta(n_{22} + n_{33})]\frac{s}{1+s}$$

, where $\Omega_{1,2}$ and $\omega_{1,2}$ represent the amplitudes and frequencies of two microwave fields in the dual driving scheme, $n_{11}, n_{22}, n_{33}$ denote the populations in the $m_s = 0, -1, +1$ spin states, respectively, and $s$ is the saturation parameter, which is the optical power relative to the saturation power[36]. In this paper, we choose $s = 0.026$ based on the experiment and the Rabi frequencies are expressed in units of $5.0 \times 10^6$ rad/Hz. The microwave frequencies are set to be arbitrary values $\omega_1$ and $\omega_2$, but the Zeeman splitting between the $m_s = \pm 1$ spin states are considered. In the case of symmetric driving, the two Rabi frequencies are equal, i.e., $\Omega_1 = \Omega_2 = \Omega_R$.

The ODMR characteristics are evaluated by analyzing contrast and linewidth derived from the PL rate. The contrast $C$ is defined by

$$C = \frac{R(0,0,0,0,s) - R(\Omega_R, \Omega_R, \omega_{12}, \omega_{13}, s)}{R(0,0,\omega_{12}, \omega_{13}, s)}$$

, while the linewidth $\Delta\omega$ is obtained by numerically assessing spectral broadening from the PL rate, defined as the full width at half maximum (FWHM)[36]. Finally, the sensitivity is calculated as

$$S \propto \frac{\Delta\omega}{C\sqrt{R_0}}$$

, where $R_0$ represents the PL rate without applying the microwave fields[36].

## Data availability

Data used in this study is available upon request to the corresponding author.

## Acknowledgements

The authors acknowledge support from the National Research Foundation of Korea (NRF) grant (NRF-2022M3K4A1094777), Information & communication Technology Planning & Evaluation (IITP) grant (No. RS-2023-00230717) and Information Technology Research Center (ITRC) support program (IITP-2024-2020-0-01606) funded by the Korea government (MSIT), and Korea Institute of Science and Technology (KIST) Institutional Program (2E32970 and 2E32971). N. M. acknowledges support from the National Research Foundation of Korea (NRF) grant (NRF-2022R1F1A1065365) funded by the Korea government (MSIT), and the grant (RS-2023-0028535) funded by the Ministry of Education (MoE).

## Author contributions

J.Y., J.J., C.K., and D.L. conceived and designed the experiments; J.Y., and J.J. prepared the samples; J.Y., J.J., H. J., J. J. performed the wide-field quantum microscopy measurements and magnetic simulations; Y. L. conducted scanning magnetometry measurements; N.M. carried out numerical calculation on the rate equations; D.L. directed and supervised the project; J.Y., J.J., N.M., and D.L. wrote the manuscript with contributions from all co-authors.

## Competing interests

J.Y., J.J., and D.L are preparing a patent partially based on this work. The other authors declare no competing interests.

**Figure captions**

**Fig.1** Working principle of magnetic steganography. (a) The wide-field quantum microscope consists of a diamond plate with an ensemble of NV centers located about 15 nm below from the top surface. Underlying the diamond, an Omega-shaped microwave circuit generates microwave field around 2.9 GHz to drive the NV spin transitions. Magnetic image is obtained by exciting the NV centers using a green laser and collecting their fluorescence using a sCMOS camera, while optical image is obtained using a separate red LED light. Made from combinations of magnetic, Ni, and non-magnetic, Au, materials, a pixel art sample is fabricated and placed on top of the diamond plate. (b) An example of NV's ODMR measurement. Non-zero magnetic field along the NV axis induces Zeeman splitting of the spin resonances, allowing us to determine the magnitude of magnetic field. (c) Experimentally measured optical and magnetic field images on the pixel art sample in (a). The square dots are 1 x 1 $\mu m^2$ in size

and are spaced by 2 μm. While the optical image shows the overall sample structure, e.g., heart, a hidden magnetic pattern, e.g., diamond, is revealed only in the magnetic image. All scale bars = 10 μm.

**Fig.2** Magnetic steganography on barcodes and QR codes. (a) Two different magnetic barcode samples are fabricated with Ni and Au rectangle bars. The bar width and spacing are either 2 or 4 μm; the length is 20 μm. The white dashed rectangles indicate Au bars. Optically identical patterns show distinct magnetic images in the magnetic field measurements. The magnetic patterns match with the designed secret messages of "NV" and "KR". (b) Two different magnetic QR code samples are constructed with Ni and Au square dots. The dots are 2 x 2 μm$^2$ in size and are separated by 1 μm. Different QR patterns shown in the magnetic field images are invisible in the optical images. The reconstructed QR codes correspond to the homepage of Korea University (http://www.korea.ac.kr) and the authors' laboratory (http://www.qdl.korea.ac.kr). All scale bars = 10 μm.

**Fig.3** Comparison of three magnetic imaging modes. (a) A schematic of the ODMR measurement. The NV's photoluminescence signal is plotted as a function of microwave frequency. Four pairs of resonances result from the NV centers with different crystal axes in a diamond: [111], [$\bar{1}$11], [1$\bar{1}$1], and [11$\bar{1}$]. Upon an applied magnetic field of ~ 400 G, the pairs are separated, and we focus on one pair of the resonances, e.g., [111], for the imaging experiments (dashed boxes). Magnetic information can be obtained from either frequency shift, change in linewidth or contrast of the resonance peak. While the frequency shift gives quantitative magnitude and sign of the magnetic field, the linewidth and contrast provide only qualitative information about the magnetic sample. However, steganography requires imaging of the hidden structure rather than quantitative information about the sample, suggesting the latter two modes are sufficient for the image steganography applications. (b) Measured magnetic images from the pixel art, barcode, and QR code samples using three imaging modes: frequency shift, linewidth, and contrast mapping. More detail comparison on the modes is available in the main text. All scale bars = 10 μm.

**Fig.4** Magnetic simulation on three imaging modes. (a)-(b) Micromagnetic simulation (a) and scanning magnetometry measurement (b) based on a single NV center. The OOMMF simulation (50×50×50 nm³ mesh size) produces the three-dimensional distribution of magnetic stray field around square Ni dots (1 x 1 µm², 2 µm separation) at 1 µm above the sample. The stray field is then projected onto the NV's axis followed by emulating the ODMR data: frequency shift, linewidth, and contrast variations in every pixel of the image. The inset images show simulations of a single Ni dot. The lower graphs display the linecut data corresponding to the dashed lines in the images. (b) The scanning experiment is conducted at the height of ~1 µm from the surface over a corner section of the QR code sample. (c) Micromagnetic simulation based on the wide-field quantum microscope configuration in this study. In addition to the simulation procedure used for the single NV center in (a), we incorporate the field averaging effect resulting from the several NV centers in the diffraction-limited laser spot with ~ 100 µm in diameter. The largest changes in frequency shift and linewidth occur at the dot margins, which can complicate image analysis particularly in cases of closely spaced dots. On the other hand, the contrast change is more focused at the dot center, implying advantages in the image recognition. All scale bars = 5 µm.

**Fig.5** Dual driving measurement for fast imaging. (a) A schematic of the NV's qutrit states. The degenerated $m_s = -1$ and $m_s = +1$ states can be split by an external magnetic field and the microwave at two different frequencies can be applied to independently drive the qubit transitions between $m_s = 0$ and $m_s = -1$, and $m_s = 0$ and $m_s = +1$. The blue dotted and red dashed arrows indicate single driving of the qubit transition, while the black solid arrows for simultaneous driving of the qutrit transitions, i.e., dual driving. (b) Comparison of the measured QR code images obtained by the single driving and dual driving. The Pearson correlation coefficient is plotted as a function of the imaging time, which measures the relative quality of an image taken at a given time against the final image after long measurement time. (c) The QR code images acquired from the dual (① and ③) and single driving (② and ④) taken at two different measurement times, i.e., 100 and 1,000 seconds. The single driving takes almost three times longer imaging to reach the same Pearson correlation coefficient of ~ 0.65 at ①.

**Fig.6** Numerical calculation for dual driving. (a) A schematic of the NV's 7 energy levels: 3 ground states and 3 excited states for spin triplets, and 1 shelving state for spin singlet. $\Gamma_1$, $\Gamma_4$, $\Gamma_5$, $\Gamma_6$, and $\Gamma_p$ denote the excitation and relation rates, and $\Omega_R$ represents the Rabi frequency between the ground spin states. A hypothetical state 8 is used for the electrons at the ground states to be excited and then to be relaxed to the excited states. (b) Calculated ODMR spectrum using single and dual driving at three different Rabi frequencies, $\Omega_R$. The Rabi frequencies are expressed in units of $5.0 \times 10^6$ rad/Hz. (c) Calculated ODMR contrast $C$ and linewidth $\Delta\omega$, and magnetic field sensitivity $S$ are plotted as a function of $\Omega_R$ based on the single and dual driving schemes. The difference in $C$ is much larger than that of $\Delta\omega$, providing the overall enhancement in $S$ for the dual driving and about twofold enhancement is realized at $\Omega_R = 1$.

**Fig. 1**

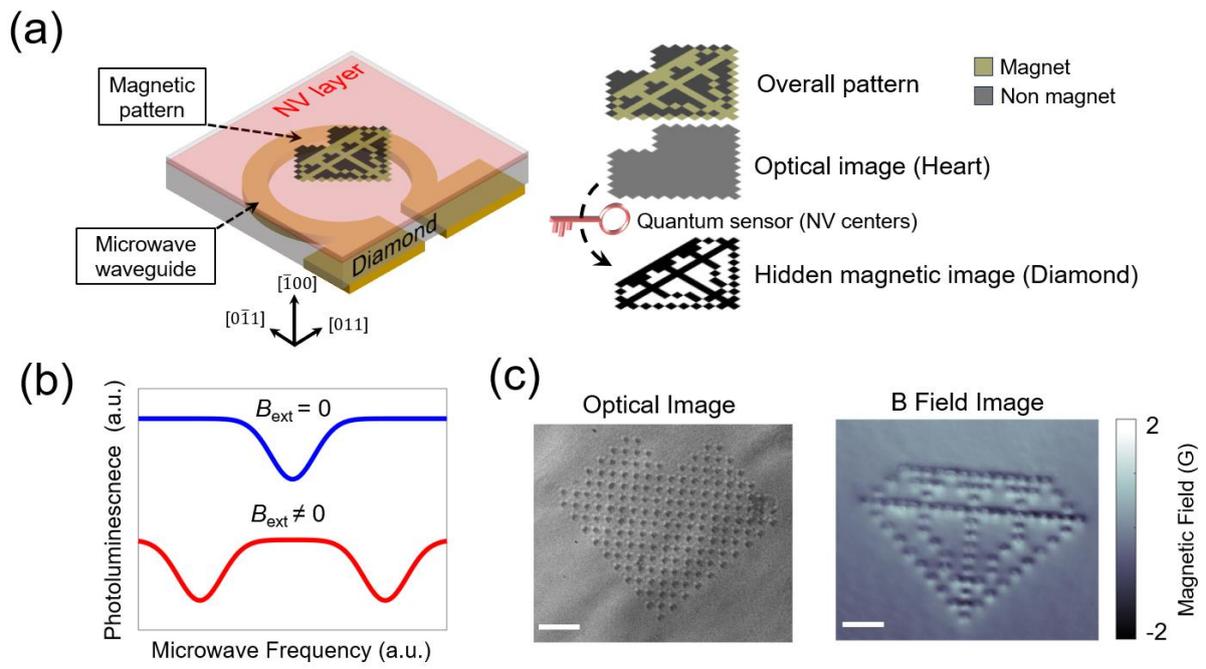

**Fig. 2**

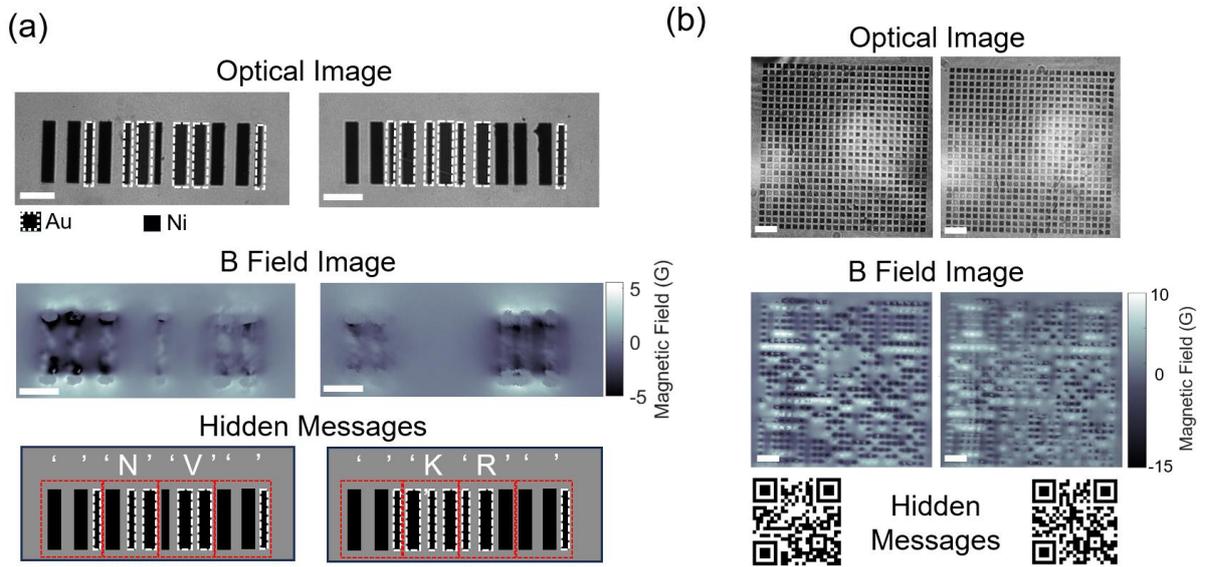

**Fig. 3**

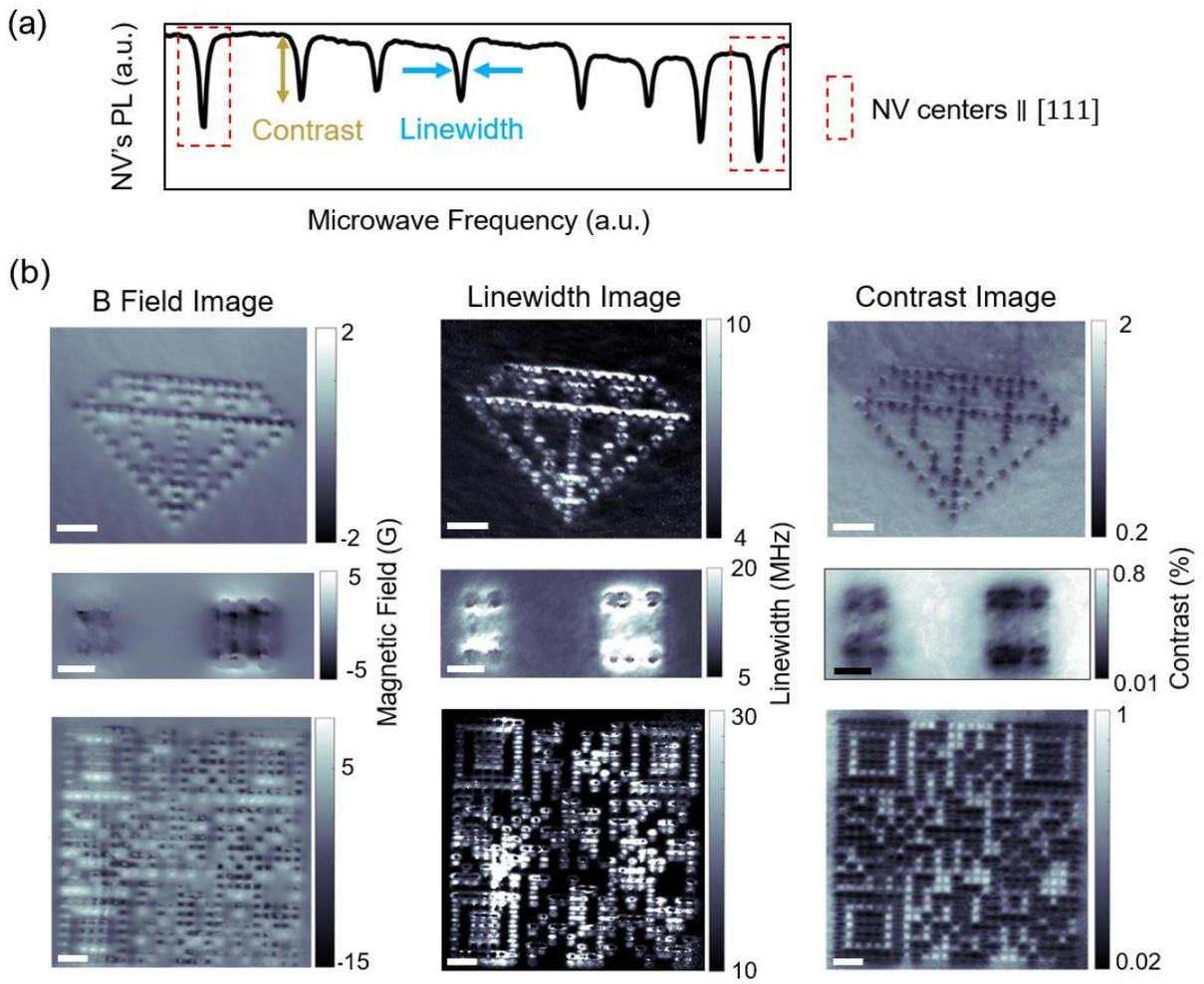

**Fig. 4**

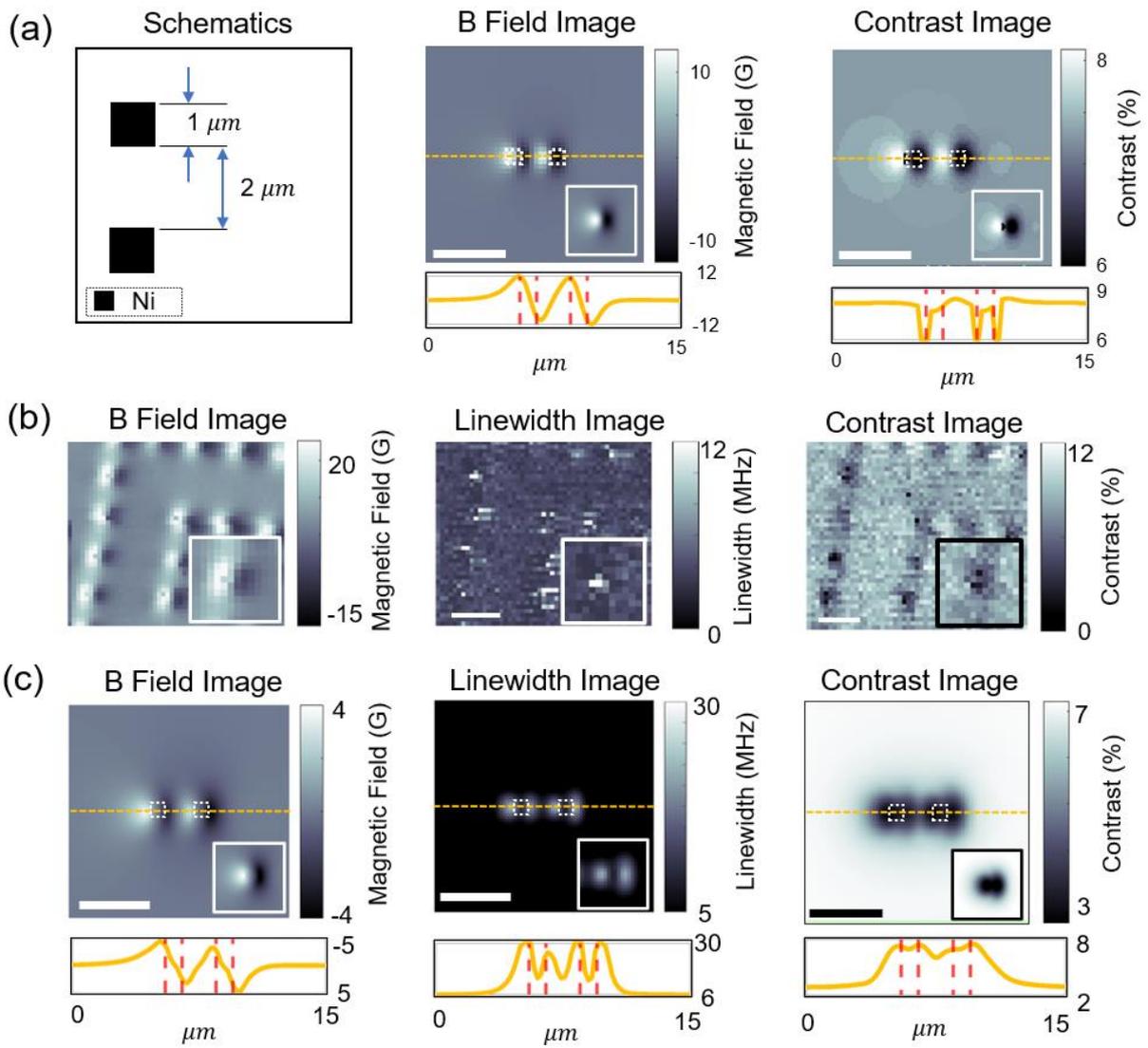

**Fig. 5**

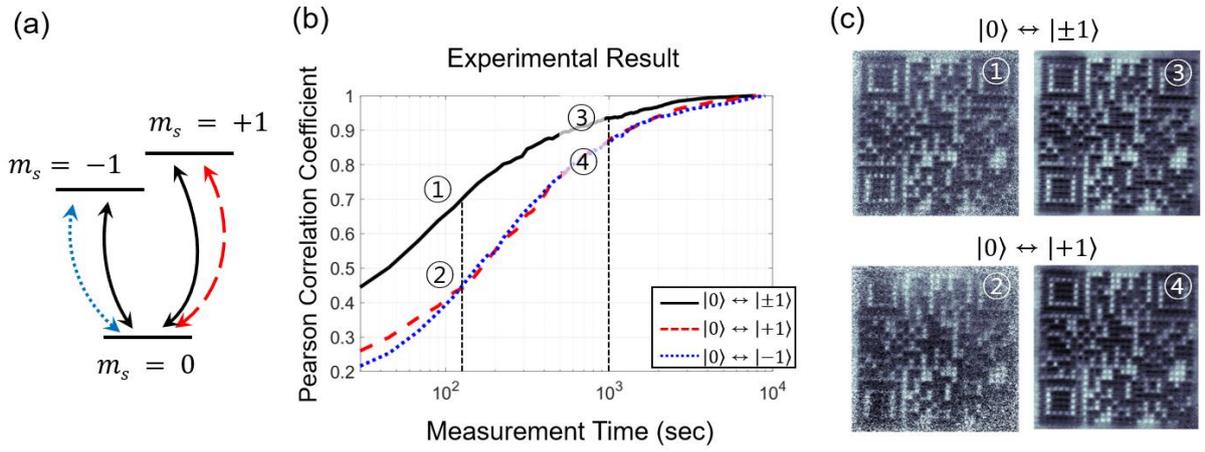

**Fig. 6**

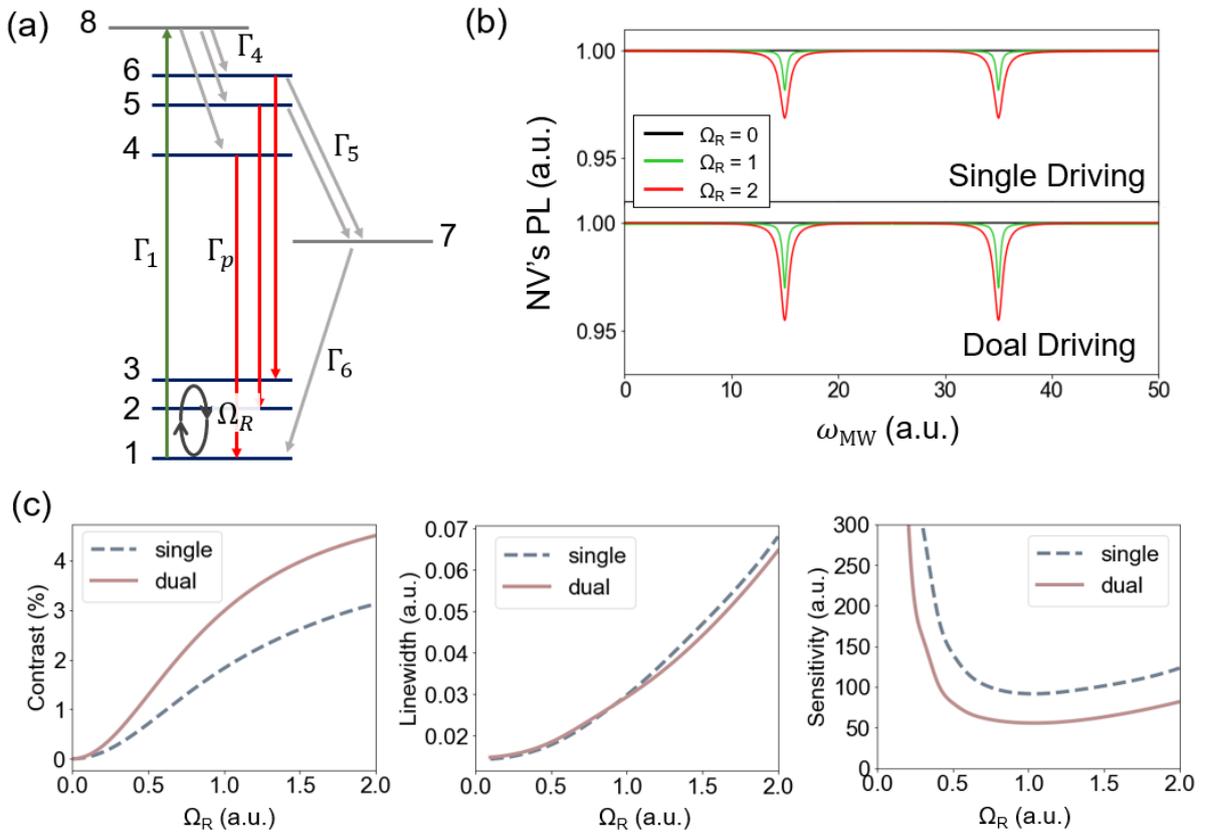